\documentclass[a4paper, 12pt, oneside]{article}
\usepackage{amsmath}
\usepackage{amssymb}
\usepackage{enumerate}
\usepackage{graphicx}
\usepackage{epsf}
\usepackage{verbatim}
\usepackage{tikz}
\usepackage{systeme,mathtools}
\usetikzlibrary{matrix,arrows,decorations.pathmorphing}

\usepackage[amsmath,thmmarks]{ntheorem}
\newcounter{thm} 

\newtheorem{Thm}[thm]{Theorem}

\newtheorem{Cor}[thm]{Corollary}

\theoremstyle{nonumberplain}
\theoremheaderfont{\normalfont\itshape}
\theorembodyfont{\normalfont}
\theoremseparator{.}
\theoremsymbol{\ensuremath{\square}}
\newtheorem{proof}{Proof}

\def \Z {\mathbb Z}

\def \C {\mathbb C}

\def \Q {\mathbb Q}

\def \WW {\mathbf{W}}
\def \i {\mathfrak{i}}

\def \d {\mathrm{d}}
\def \tr {\mathrm{Tr}}

\def \h {\hbar}

\def \D {\mathcal{D}}

\def \W {\mathcal{W}}

\def \L {\mathcal{L}}
\def \M {\mathcal{M}}
\def \Q {\mathcal{Q}}

\def \OE {\operatorname{OE}}
\def \Wa {\widehat{\alpha}}

\begin{document}
	\begin{titlepage}
		
		\title{The ordered exponential representation of GKM using the $W_{1+\infty}$ operator}
		\author{Gehao Wang}
		\date{}
		\maketitle

		\begin{abstract}
		The generalized Kontsevich model (GKM) is a one-matrix model with arbitrary potential. Its partition function belongs to the KP hierarchy. When the potential is monomial, it is an $r$-reduced tau-function that governs the $r$-spin intersection numbers. In this paper, we present an ordered exponential representation of monomial GKM in terms of the $W_{1+\infty}$ operators that preserves the KP integrability. In fact, this representation is naturally the solution of a $W_{1+\infty}$ constraint that uniquely determines the tau-function. Furthermore, we show that, for the cases of Kontsevich-Witten and generalized BGW tau-functions, their $W_{1+\infty}$ representations can be reduced to their cut-and-join representations under the reduction of the even time independence and Virasoro constraints.
		\end{abstract}
		\vspace{20pt}
		
		
	\end{titlepage}

\section{Introduction}

For any potential $V(z)$, the generalized Kontsevich model (GKM) is described by the following matrix integral over the space of Hermitian matrices (see, e.g., \cite{KMMM},\cite{KMMMZ},\cite{KMMMZ2},\cite{M1},\cite{Agkm})
\begin{equation*}
	Z_U(\Lambda)=\mathcal{C}^{-1}\int [\d\Phi] \exp\left(-\tr (V(\Phi)-\Phi V^{'}(\Lambda)) \right),
\end{equation*}
where $\Lambda$ is the diagonal matrix, the label $U=V^{(2)}(z)$ and 
\begin{equation*}
	\mathcal{C}=\frac{\Delta(\lambda)}{\Delta(V^{'}(\lambda))}\sqrt{\det\left(\frac{2\pi}{V^{(2)}(\lambda)}\right)}\exp\left(\tr(\Lambda V^{'}(\Lambda)-V(\Lambda))\right).
\end{equation*}
It generalizes the well-known Kontsevich matrix model \cite{K}, and describes a universal non-perturbative partition function of topological string models. When the potential is monomial, that is, $U(z)=z^{n}$, the partition function $\tau_n$ of GKM is an $r$-reduced KP tau-function subject to the $r$-spin intersection theory \cite{EW} with $r=n+1$. 

Ward identities usually provide a way to generate Virasoro and $W$-constraints for  matrix models (see, e.g., \cite{M1}). Performing some changes of variables in the matrix integral can lead to a system of constraints that determine the partition function. In a recent series of study (e.g., \cite{MM2}, \cite{MMM} and \cite{MMM2}), the $W$-constraints of GKM are revisited after the first investigation in \cite{Z}. It turns out that one can form a single constraint that uniquely determines the partition function of the monomial GKM by combining all the $W$-constraints in a certain way. This single equation contains the grading operator. Hence it naturally implies an ordered exponential representation for $\tau_n$ using the $W$-operators, called $W$-representations. In the case $n=1$, it is the cut-and-join representation (Eq.\eqref{CJKW}) of the Kontsevich-Witten tau-function presented in \cite{A1}. However, these $W$-operators generally do not belong to the $W_{1+\infty}$ algebra that describes a fundamental symmetry for the KP hierarchy \cite{FKN}.

Let us remind ourselves that the ordered exponential $\operatorname{OE}[a](\h)$ with the time parameter $\h$, also called the time-ordered (or path-ordered) exponential, is the unique solution of the equation
\begin{equation*}
	\frac{\d}{\d \h}\operatorname{OE}[a](\h)=a(\h)\operatorname{OE}[a](\h)
\end{equation*}
with the initial condition $\operatorname{OE}[a](0)=1$. In our context, we will always use $\h$ as the time parameter, and write $\operatorname{OE}[a]$ for simplicity. If we let $e^W=\operatorname{OE}[a]$, then the expression of $W$ can be computed from $a$ using the Magnus expansion \cite{M}. When the elements in $a$ are commutative,  $\operatorname{OE}[a]=e^A$, where $\partial_{\h}A=a$.
  
Inspired by the work in \cite{MMM}, \cite{MMM2} and \cite{M2}, in this paper, we will first present a $W_{1+\infty}$ constraint that not only uniquely determines the monomial GKM, but also implies an ordered exponential representation that preserves the KP integrability. We aim to study the relations between the $W_{1+\infty}$ and  $W$-constraints of GKM. 

Let $\D$ be the Lie algebra of differential operators on the circle
\begin{equation*}
	z^n\partial_z^m, \mbox{ where } n,m\in\Z,m\geq 0.
\end{equation*}
The algebra $W_{1+\infty}$ of fermion bilinear operators is a central extension of $\D$. We can express the $W_{1+\infty}$ symmetries of tau-functions using the corresponding one-body operators in $\D$. The standard basis for $W_{1+\infty}$ are the $W^{(k)}$-algebras indexed by the spin $k$, (see, e.g., \cite{PRS}). For example, under the boson-fermion correspondence, the Heisenberg operators $\widehat{\alpha}_n$ corresponds to $-z^n$ in the $W^{(1)}$-algebra, where
\[
\widehat{\alpha}_n=
\begin{cases}
	q_{-n} & n<0 \\
	n\frac{\partial}{\partial q_n} &  n>0.
\end{cases}
\]
The Virasoro operators $\widehat{L}_n$ correspond to $-z^{1+n}\partial_z-\frac{n+1}{2}z^n$ in the $W^{(2)}$-algebra, where
\begin{align}\label{hatL}
	\widehat{L}_n
	&=\sum_{i>0,i+n>0} (i+n)q_i\frac{\partial}{\partial q_{i+n}}+\frac{1}{2}\sum_{i+j=n} ij\frac{\partial^2}{\partial q_i\partial q_j}+\frac{1}{2}\sum_{i,j>0,i+j=-n}q_iq_j.
\end{align}
The operators $\widehat{M}_n$ defined in Eq.\eqref{hatM} corresponds to the elements in $W^{(3)}$-algebra. In particular, the operator $\widehat{M}_0$ is in the cut-and-join representation for the Hurwitz tau-function \cite{GJ}.

For $R\in\D$, let $W_R$ be the fermionic operator corresponding to $R$, and let $\widehat{\WW}_R$ be the bosonic $W_{1+\infty}$ operator $\widehat{W}_R$ with the extra scaling $q_k\rightarrow \h^kq_k$. Also, we denote $\tau(\h)$ to be the tau-function $\tau$ with the scaling $q_k\rightarrow \h^kq_k$. Let us present the $W_{1+\infty}$ representation of $\tau_n(\h)$ as follows.
\begin{Thm}\label{main}
	For the tau-function $\tau_n(\h)$ of the monomial GKM,  we have
	\begin{equation}\label{taun}
		\tau_n(\h)=\OE[\h^{-1}\widehat{\WW}_{R_n}]\cdot 1,
	\end{equation} 
where, for $D=z\partial_z$,
\begin{equation}\label{Rn}
	R_n=-D-\frac{1}{n+1}\left\{z-z^{-n-1}\left(D-\frac{n}{2}\right)\right\}\left\{ \left(z-z^{-n-1}(D-\frac{n}{2}) \right)^{n+1} -z^{n+1}\right\}.
\end{equation}
Equivalently, it is uniquely determined by the single $W_{1+\infty}$ constraint
\begin{equation}\label{Wns}
	\left(\widehat{L}_0-\widehat{\WW}_{R_n}\right)\cdot \tau_{n}(\h)=0.
\end{equation}
\end{Thm}
According to the results in \cite{Agkm},  all GKMs with polynomial potentials of the same degree are equivalent to each other up-to the conjugation of Heisenberg and Virasoro operators. Since the conjugation of Heisenberg operators sends elements in $W^{(n)}$-algebra to the ones in $W^{(n-1)}$-algebra, and the conjugation of Virasoro operators stabilizes the $W^{(n)}$-algebra, we can use Eq.\eqref{Wns} to obtain equivalent descriptions of all GKMs with polynomial potentials using elements in the $W^{(i)}$-algebras, where $1\leq i\leq n+3$. Examples about the conjugation of Heisenberg-Virasoro operators on constraints can be found in other articles such as \cite{GW} and \cite{A6}. 

For the operator $Q_n$ belonging to the $W^{(4)}$-algebra, the bosonic version is defined in Eq.\eqref{hatQ}. We give the explicit bosonic expression of Eq.\eqref{taun} in the case of the Kontsevich-Witten tau-function that governs the intersection theory on the moduli spaces of closed Riemann surfaces.
\begin{Thm}\label{WKW}
	The Kontsevich-Witten tau-function $\tau_{KW}(\h)$ can be represented as the following ordered exponential of $W_{1+\infty}$ operators
	\begin{equation}\label{OEKW}
		\tau_{KW}(\h)=\OE\left[\h^{-1}\widehat{\WW}_{R_1} \right]\cdot 1.
	\end{equation}
	where
\begin{equation}\label{WKWo}
	\widehat{\WW}_{R_1}=\frac{3\h^3}{2}\widehat{M}_{-3}+\frac{\h^3}{8}q_3-\frac{\h^6}{2}\widehat{Q}_{-6}-\frac{21\h^6}{40}\widehat{L}_6.
\end{equation}
	Equivalently, it is uniquely determined by the single $W_{1+\infty}$ constraint
	\begin{equation}\label{W1s}
		\left(\widehat{L}_0-\widehat{\WW}_{R_1}\right)\cdot \tau_{KW}(\h)=0.
	\end{equation}
\end{Thm}

We also present an explicit bosonic $W_{1+\infty}$ representation of the generalized Brezin-Gross-Witten (BGW) tau-function $\tau_N(\h)$ (see, e.g., \cite{MMS} and \cite{Abgw}) as follows.
\begin{Thm}\label{WN}
	The generalized BGW-tau function $\tau_N(\h)$ can be represented as the following ordered exponential of $W_{1+\infty}$ operators
	\begin{equation}\label{OEN}
		\tau_N(\h)=\OE\left[\h^{-1} \widehat{\WW}_{R_N}\right]\cdot 1.
	\end{equation}
where
\begin{equation}\label{WNo}
	\widehat{\WW}_{R_N}=\frac{3\h}{4}\widehat{M}_{-1}+\frac{\h}{4}\left(\frac{1}{4}-N^2\right)q_1-\frac{\h^2}{8}\widehat{Q}_{-2}-\frac{\h^2}{8}\left(\frac{1}{4}-N^2\right)\widehat{L}_{-2}.
\end{equation}
	Equivalently, it is uniquely determined by the single $W_{1+\infty}$ constraint
	\begin{equation}\label{WNs}
		\left(\widehat{L}_0-\widehat{\WW}_{R_N}\right)\cdot \tau_N(\h)=0.
	\end{equation}	
\end{Thm}

For the cases of $\tau_{KW}$ and $\tau_N$, we will demonstrate how to reduce their $W_{1+\infty}$ representation to their $W$-representations, formerly known as the cut-and-join representations.  
\begin{Cor}\label{CJ}
	Under the Virasoro constraints and KdV reduction, the ordered exponential representations \eqref{OEKW} and \eqref{OEN} are equivalent to the cut-and-join representations of $\tau_{KW}$ and $\tau_N$ introduced in \cite{A1} and \cite{Abgw}, respectively. 
\end{Cor}
The proof of this corollary is an evidence that, the bosonic $W_{1+\infty}$ constraint and the $W$-constraint of GKM differ by some Virasoro and other lower constraints. In other words, by adding these extra parts to the $W$-representation, we can recover the KP integrability. 

The paper is organized as follows. In Section \ref{S2}, we briefly review the bosonic and fermionic realization of the $W_{1+\infty}$ algebra. In Section \ref{S3}, we remind ourselves about the Sato Grassmannian description of the KP hierarchy and Kac-Schwarz operators. In Section \ref{S4}, we will prove our presented theorems. Finally, in Section \ref{S5}, we recall the results about the cut-and-join representations for $\tau_{KW}$ and $\tau_N$, and prove Corollary \ref{CJ}.

\section{The $W_{1+\infty}$ algebra}\label{S2}
For more details about the $W_{1+\infty}$ algebra, we refer to the article \cite{FKN} and references therein.

For the two fermionic fields
\begin{equation*}
	\psi(z)=\sum_{n\in\Z}\psi_nz^{-n}\quad\mbox{ and }\quad \psi^{*}(z)=\sum_{n\in\Z}\psi_n^{*}z^{-n-1},
\end{equation*} 
the current operator $\alpha_n$ is defined as
\begin{equation*}
	\alpha_n=\sum_{j\in\Z}:\psi_{-j}\psi_{j+n}^*:\quad \mbox{ with }\quad [\alpha_m,\alpha_n]=m\delta_{m+n,0},
\end{equation*}
which are included in the generating series
\begin{equation*}
	\alpha(z)=:\psi(z)\psi^{*}(z):=\sum_{n\in\Z}\alpha_nz^{-n-1}.
\end{equation*}
The normal product $:\psi_m\psi_n^*:$ of a quadratic monomial in fermions is defined as
\begin{equation*}
	:\psi_m\psi_n^*:
	=\psi_m\psi_n^*-\langle \psi_m\psi_n^*\rangle=
	\begin{cases}
		\psi_m\psi_n^* & \mbox{if }  m\leq 0 \mbox{ or if } n> 0;\\
		-\psi_n^*\psi_m & \mbox{if } m> 0 \mbox{ or if } n\leq 0.
	\end{cases}
\end{equation*}

Suppose $a$ is an operator in $\D$. We define the fermion bilinear operator $W_a$ as
\begin{equation*}
	W_a=\text{res}_z\left(:\psi^{*}(z)a\psi(z):\right).
\end{equation*}
They generate the algebra $W_{1+\infty}$, which is the central extension of $\D$. As mentioned in \cite{FKN}, for $m\geq 0, m\in\Z$, the following field
\begin{align}
	W^{(m+1)}(z)&=\sum_{n\in\Z}W_n^{(m+1)}z^{-n-m-1}\nonumber\\
	&=\frac{m!}{2^m(2m-1)!!}\sum_{j=0}^{m}(-1)^j\binom{m}{j}^2:\left(\partial_z^{m-j}\psi(z)\right)\left(\partial_z^j\psi^*(z)\right): \label{Wfield}
\end{align}
describes the standard basis of the algebra $W_{1+\infty}$. We define $a^*\in\D$ to be the adjoint operator of $a$, if $a^*$ satisfies
\begin{equation*}
	\text{res}_z\left(f(z)(a\cdot g(z))\right)=\text{res}_z\left(g(z)(a^*\cdot f(z))\right).
\end{equation*}
For example, 
\begin{equation}\label{adex}
	(z^n)^*=z^n,  \quad (z^n\partial_z^k)^*=(-\partial_z)^kz^n,\quad \mbox{and}\quad(z^nD^k)^*=z^{-1}(-D)^kz^{n+1}.
\end{equation}
Using the adjoint operators, we can re-write Eq.\eqref{Wfield} as
\begin{align*}
	W_n^{(m+1)}&=\text{res}_z\left\{\frac{m!}{2^m(2m-1)!!}\sum_{j=0}^{m}(-1)^j\binom{m}{j}^2:z^{m+n}\left(\partial_z^{m-j}\psi(z)\right)\left(\partial_z^j\psi^*(z)\right): \right\}\\
	&=\text{res}_z\left\{:\psi^{*}(z)\left(-\frac{m!}{2^m(2m-1)!!}\sum_{j=0}^{m}\binom{m}{j}^2\partial_z^jz^{m+n}\partial_z^{m-j}\right)\psi(z):\right\}.
\end{align*}
Therefore, we have
\begin{equation*}
	W^{(1)}(z)=\sum_{n\in\Z}\alpha_nz^{-n-1}\quad\mbox{ and }\quad
	W^{(2)}(z)=\sum_{n\in\Z}L_nz^{-n-2},
\end{equation*}
where the nodes 
\begin{equation}\label{nodes1}
	\alpha_n=W_{-z^n} \quad\mbox{ and }\quad L_n=W_{-z^{1+n}\partial_z-\frac{n+1}{2}z^n}
\end{equation}
span the Heisenberg and Virasoro algebra respectively. Furthermore, we have
\begin{equation*}
	W^{(3)}(z)=\sum_{n\in\Z}M_nz^{-n-3},
	\quad\mbox{ and }\quad W^{(4)}(z)=\sum_{n\in\Z}Q_nz^{-n-4},
\end{equation*}
where the nodes $M_n$ and $Q_n$ are in the form
\begin{align}
	&M_n=W_{-z^{2+n}\partial^2_z-(n+2)z^{1+n}\partial_z-\frac{1}{6}(n+1)(n+2)z^n}.\label{nodes2}\\
	&Q_n=W_{-z^{n+3}\partial_z^3-\frac{3}{2}(n+3)z^{n+2}\partial_z^2-\frac{3}{5}(n+2)(n+3)z^{n+1}\partial_z-\frac{1}{20}(n+1)(n+2)(n+3)z^n}\label{nodes3}.
\end{align}

On the other hand, since
\begin{equation*}
	:\partial_z^k\psi(z)\partial_z^l\psi^*(z):=\sum_{i=k+1}^{k+l+1}\frac{(-1)^{i-1-k}}{i}\binom{l}{i-1-k}\partial_z^{k+l+1-i}P^{(i)}(z),
\end{equation*}
where
\begin{equation*}
	P^{(i)}(z)=:e^{-\phi(z)}\partial_z^ie^{\phi(z)}:\quad\mbox{and}\quad \partial_z\phi(z)=\alpha(z),
\end{equation*}
we can obtain a bosonic realization of the $W_{1+\infty}$ algebra as
\begin{align}
	W^{(m+1)}(z)&=\sum_{n\in\Z}W_n^{(m+1)}z^{-n-m-1}\nonumber\\
	&=\sum_{l=0}^m\frac{(-1)^{l}}{(m+1-l)l!}\left(\frac{m!}{(m-l)!}\right)^2\frac{(2m-l)!}{(2m)!}\partial_z^{l}P^{(m+1-l)}(z) \label{field}.
\end{align}
For example:
\begin{align*}
		W^{(1)}(z)&=\alpha(z),\quad W^{(2)}(z)=\frac{1}{2}:\alpha(z)^2:,\\
		W^{(3)}(z)&=\frac{1}{3}:\alpha(z)^3:,\\
		W^{(4)}(z)&=\frac{1}{4}:\alpha(z)^4:+\frac{1}{10}:\alpha(z)\partial_z^2\alpha(z):-\frac{3}{20}:\left(\partial_z\alpha(z)\right)^2:
\end{align*}

In terms of the differential operators on bosonic Fock space in variables $q_k$, we will write the bosonic $W_{1+\infty}$ operators corresponding to $W_a$ as $\widehat{W}_a$. 
The operators $\widehat{M}_n$ can be expressed as
\begin{align}\label{hatM}
	\widehat{M}_n&=\frac{1}{3}\sum_{a+b+c=n}:\widehat{\alpha}_a\widehat{\alpha}_b\widehat{\alpha}_c:\nonumber\\
	&=\sum_{\substack{i,j>0\\i+j+n>0}}(i+j+n)q_iq_j\frac{\partial}{\partial q_{i+j+n}}+\sum_{\substack{i,j>0\\i+j-n>0}}ijq_{i+j-n}\frac{\partial^2}{\partial q_i\partial q_j}\nonumber\\
	&\quad\quad+\frac{1}{3}\sum_{\substack{i,j>0\\n-i-j>0}}ij(n-i-j)\frac{\partial^3}{\partial q_i\partial q_j\partial q_{n-i-j}}
	+\frac{1}{3}\sum_{\substack{i,j>0\\-n-i-j>0}}q_iq_jq_{-n-i-j}.
\end{align}
The operators $\widehat{Q}_n$ are 
\begin{align}\label{hatQ}
	\widehat{Q}_n&=\frac{1}{4}\sum_{a+b+c+d=n}:\widehat{\alpha}_a\widehat{\alpha}_b\widehat{\alpha}_c\widehat{\alpha}_d:-\frac{1}{4}\sum_{a+b=n}(a+1)(b+1):\widehat{\alpha}_a\widehat{\alpha}_b:\nonumber\\
	&\quad\quad+\frac{1}{10}(n+2)(n+3)\widehat{L}_n.
\end{align}
In general, Eq.\eqref{field} can be used to determine the operators $\widehat{W}_a$ for any $a\in\D$.

\section{Kac-Schwarz operators and tau-functions of the KP hierarchy}\label{S3}
Let us review the Sato Grassmannian description \cite{S} of the tau-function $\tau$ of the KP hierarchy. It is well-know that these tau-functions form the orbit of the vacuum state $|0\rangle=\psi_0^*\psi_1^*\psi_2^* \dots |\infty\rangle$ under the action of the group $GL(\infty)$ associated with the Lie algebra $\widehat{\mathfrak{gl}(\infty)}$ (see, e.g.,\cite{DJKM} and \cite{MJD}). In fermionic representation, there exists a group element $\widetilde{G}$ with positive energy in the form
\begin{equation}\label{tildeG}
	\widetilde{G}=\exp\left(\sum_{m\geq 0,n\geq 1}A_{m,n}\psi_{-m}\psi_{-n}^*\right),
\end{equation}
such that the tau-function $\tau$ can be expressed using a set of fermionic operators $\{\widetilde{\psi}_{n}\}$ as 
\begin{equation*}
	\tau=\widetilde{G} |0\rangle= \widetilde{\psi}_{0}\widetilde{\psi}_{1}\widetilde{\psi}_{2}\dots  |\infty\rangle,
\end{equation*}
where, for $m=0,1,2,\dots$,
\begin{equation*}
	\widetilde{\psi}_{m}=\widetilde{G}\psi_{m}^*\widetilde{G}^{-1}=\psi_{m}^*-\sum_{n=1}^{\infty} A_{m,n}\psi_{-n}^*.
\end{equation*}
These operators correspond to the canonical basis vectors $\{\Psi_{m+1}(z)\}$ for the tau-function $\tau$, where
\begin{equation*}
	\Psi_{m+1}(z)=z^m-\sum_{n=1}^{\infty} A_{m,n}z^{-n},
\end{equation*}
and they generate the point of the Sato Grassmannian $\W$ corresponding to the tau-function $\tau$. The set of vectors $\{z^m\}$ form a basis for the trivial tau-function $1$. 

Using the boson-fermion correspondence and Wick's theorem, we can write the tau-function $\tau$ in the form of correlation function
\begin{equation*}
	\tau\left(-[Z^{-1}]\right)=\frac{\det_{i,j=1}^M\Psi_j(z_i)}{\Delta(z)}
\end{equation*}
for some integer $M$, where $\Delta(z)=\prod_{i<j}(z_j-z_i)$ is the Vandermonde determinant. The notation $\tau\left(-[Z^{-1}]\right)$ means that our tau-function is in the following Miwa parametrization for the matrix model,
\begin{equation*}
	\tau\left(-[Z^{-1}]\right)=\tau\left(q_{k}=-\tr \Lambda^{-k}\right).
\end{equation*} 

In the next section, the Miwa parametrization used to describe the tau-functions is $q_{k}=\tr \Lambda^{-k}$. In this case, for the same group element $\widetilde{G}$, the canonical basis vectors $\{\Psi^{\perp}_{m+1}(z)\}$ for the tau-function $\tau\left([Z^{-1}]\right)$ are in the form 
\begin{equation*}
	\Psi^{\perp}_{m+1}(z) =\langle 0|\psi_{m+1}\psi^*(z)\widetilde{G}|0\rangle=z^{m}+\sum_{k=1}^{\infty} A_{k-1,m+1}z^{-k}.
\end{equation*}
They are usually referred as the adjoint canonical basis vectors and define the orthogonal complement $\mathcal{W}^{\perp}$ of the space $\mathcal{W}$ due to the relation:
\begin{equation}\label{or}
	\text{res}_z\left(\Psi_{i}(z)\Psi^{\perp}_{j}(z)\right)=0,  \mbox{ for }i,j\geq 1.
\end{equation}

If an operator $a\in\D$ satisfies $a\cdot\W\subset \W$, then, in the bosonic Fock space,  we have $\widehat{W}_{a}\cdot \tau=C\tau$ for some constant $C$. Such operator $a$ is usually called the Kac-Schwarz operator for the tau-function $\tau$ \cite{KacS}. In addition, the adjoint operator $a^*$ will also satisfy 
\begin{equation}\label{adjKS}
	a^*\W^{\perp}\subset \W^{\perp} \quad\mbox{and}\quad \widehat{W}_{a^*}\cdot \tau=C_1\tau
\end{equation}
for some constant $C_1$.

\section{Representation of GKM using $W_{1+\infty}$ operators}\label{S4}
Let $\D_{-}$ be the subalgebra of $\D$, generated by the differential operators $z^{-n}D^m$ with $n,m\in\Z, m\geq 0$ and $n\geq 1$. Suppose $b$ is an operator in $\D_{-}$ such that
\begin{equation*}
	\left(D+b\right)\cdot\W\subset \W.
\end{equation*}
We are particularly interested in this type of Kac-Schwarz operator, because it always uniquely determines the space $\W$. Indeed, if $\{\Phi_j(z)\}$ is a set of basis vectors of $\W$, where $\Phi_j(z)$ is in the form
\begin{equation}\label{Phi}
	\Phi_j(z)=z^{j-1}+\sum_{i>1-j} \phi_{j,i}z^{-i},\quad j\geq 1
\end{equation}
for some coefficients $\{\phi_{j,i}\}$, then
\begin{align*}
	&\left(D+b\right)\cdot\Phi_1(z)=c^{(1)}_1\Phi_1(z);\\
	&\left(D+b\right)\cdot\Phi_j(z)=c^{(j)}_1\Phi_j(z)+c^{(j)}_2\Phi_{j-1}(z)+\dots+c^{(j)}_j\Phi_1(z),
\end{align*}
for some coefficients $\{c^{(j)}_i\}$, (in fact, $c^{(j)}_1=j-1$). Starting from $\Phi_1(z)$, the operator $\left(D+b\right)$ will recursively determine all the basis vectors $\{\Phi_j(z)\}$. Since $b\in\D_{-}$, the operator $\widehat{W}_b$ has positive energy. Therefore, the function
\begin{equation*}
	\OE[\h^{-1}\widehat{\WW}_b]\cdot 1
\end{equation*}
is a well-defined tau-function of the KP hierarchy. If the tau-function $\tau$ has normalization $\tau(0)=1$ and $D+b$ is its Kac-Schwarz operator, then it satisfies
\begin{equation*}
	\left(\widehat{L}_0-\widehat{\WW}_b\right)\cdot\tau(\h)=0, \quad\mbox{which is equivalent to}\quad \left(\partial_{\h}-\h^{-1}\widehat{\WW}_b\right)\cdot\tau(\h)=0.
\end{equation*}
We can immediately conclude that
\begin{equation}\label{OEW}
	\tau(\h)=\OE[\h^{-1}\widehat{\WW}_b]\cdot 1.
\end{equation}
This kind of ordered exponential representations on a tau-function was also considered in \cite{W2}.

\subsection{The monomial GKM}\label{S41}
From the results in \cite{Agkm}, we know that the following pair of canonical Kac-Schwarz operators provide a complete description of the point of the Sato Grassmannian for the generalized Kontsevich model $Z_U$:
\begin{align*}
	K_U&=z+(zU)^{-1}D-\frac{1}{2}U^{'}U^{-2}\\
	P_U&=V^{'}\left(K_U\right)-X_U,
\end{align*}
where $X_U=V^{'}(z)$. 
For the case of the monomial potential $U=z^n$, the expressions of the corresponding operators $K$ and $X$ are 
\begin{align*}
	K_n&=z+z^{-n-1}\left(D-\frac{n}{2}\right) \\
	X_n&=\frac{z^{n+1}}{n+1}.
\end{align*}
The operator $$P_n=\frac{1}{n+1}K_n^{n+1}-X_n$$ 
is known as the quantum spectral curve operator. Furthermore, the operator $X_n$ indicates that the tau-function of the monomial GKM belongs to the $(n+1)$-reduced KP hierarchy. 
\vspace{10pt}
\noindent


\begin{proof}[of Theorem \ref{main}]
Using Eq.\eqref{adex}, let us consider the adjoint operator
\begin{equation*}
K_n^*P_n^*=\frac{1}{n+1}\left\{z-z^{-n-1}\left(D-\frac{n}{2}\right)\right\}\left\{ \left(z-z^{-n-1}(D-\frac{n}{2}) \right)^{n+1} -z^{n+1}\right\}.
\end{equation*}
The operator $P_n^*$ is in the subalgebra $\D_{-}$ with the leading term $-z^{-1}D$. Suppose 
\begin{equation*}
	W_{R_n}=-D-W_{K_n^*P_n^*}.
\end{equation*}
Then, $R_n$ is in $\D_{-}$, and Theorem \ref{main} follows from Eq. \eqref{adjKS} and \eqref{OEW}.
\end{proof}

\vspace{10pt}
\noindent
{\bf Remark:} 
\emph{In the article \cite{M2}, the author discussed the origin of a single equation that uniquely determines the GKM from the matrix-valued Ward identity and matrix Gross-Newmann equation. It would be interesting to add the $W_{1+\infty}$ constraint, and compare the three representations at the same time. We will discuss this elsewhere.}

\subsection{The Kontsevich-Witten tau-function}\label{S42}
The Kontsevich matrix model is described by the following integral 
\begin{equation*}
	Z_{KW}=\frac{\int [\d\Phi] \exp\left(-\tr (\frac{\Phi^3}{6}+\frac{\Lambda\Phi^2}{2}) \right)}{\int [\d\Phi] \exp\left(-\tr \frac{\Lambda\Phi^2}{2} \right)}.
\end{equation*}
Under the Miwa parametrization $q_{2k+1}=\tr(\Lambda^{-2k-1})$, the tau-function $\tau_{KW}$ belongs to the KdV hierarchy (\cite{K}). If we use the $W_{1+\infty}$ operators, the Virasoro constraints for $\tau_{KW}$ (see, e.g. \cite{AE},\cite{IZ}) can be expressed as  
\begin{equation}\label{VforK}
	\left(\widehat{L}_{2m}-(2m+3)\frac{\partial}{\partial q_{2m+3}}+\frac{1}{8}\delta_{m,0}\right)\cdot\tau_{KW}=0, \quad(m\geq -1).
\end{equation} 
The quantum spectral curve operator $P_1$ for $Z_K$ is in the form
\begin{equation*}
P_1=\frac{1}{2}(K_1^2-z^2)=z^{-1}D+\frac{1}{2}\left(z^{-2}(D-\frac{1}{2})\right)^2.
\end{equation*}
As mentioned in \cite{AE}, this operator generates the $\D$-module describing the tau-function \cite{DHS}.
\vspace{10pt}
\noindent

\begin{proof}[of Theorem \ref{WKW}]
For the case $n=1$, the adjoint operator $K_1^*$ and $P_1^*$ are
\begin{equation*}
	K_1^*=z-z^{-2}(D-\frac{1}{2})
\end{equation*}
and
\begin{equation*}
	P_1^*=-z^{-1}D+\frac{1}{2}\left(z^{-2}(D-\frac{1}{2})\right)^2
\end{equation*}
respectively. Let us consider 
\begin{align}\label{AlexKS}
	K_1^{*}P_1^{*}
	&=-D+\frac{3}{2}z^{-3}D^2-3z^{-3}D+\frac{5}{8}z^{-3}\nonumber\\
	&\quad\quad-\frac{1}{2}z^{-6}D^3+\frac{15}{4}z^{-6}D^2-\frac{59}{8}z^{-6}D+\frac{45}{16}z^{-6}.
\end{align}
From Eq.\eqref{nodes1} and \eqref{nodes2}, we have
\begin{align*}
	&W_{-z^{-3}D}=L_{-3}+\alpha_{-3}, \quad W_{-z^{-3}D^2}=M_{-3}+2L_{-3}+\frac{5}{3}\alpha_{-3}\\
	&W_{-z^{-6}D}=L_{-6}+\frac{5}{2}\alpha_{-6}, \quad W_{-z^{-6}D^2}=M_{-6}+5L_{-6}+\frac{55}{6}\alpha_{-6}\\
&	W_{-z^{-6}D^3}=Q_{-6}+\frac{15}{2}M_{-6}+\frac{119}{5}L_{-6}+\frac{75}{2}\alpha_{-6}.
\end{align*}
Therefore, 
\begin{equation*}
	W_{K_1^{*}P_1^{*}}=-D-\frac{3}{2}M_{-3}-\frac{1}{8}\alpha_{-3}+\frac{1}{2}Q_{-6}+\frac{21}{40}L_{-6}.
\end{equation*}
Let
\begin{equation*}
W_{R_1}=-D-W_{K_1^{*}P_1^{*}}.
\end{equation*}
Then, Theorem \ref{WKW} follows from Eq.\eqref{OEW}.
\end{proof}
\vspace{10pt}
\noindent
{\bf Remark:} 
\emph{The operator \eqref{AlexKS} is the adjoint operator of the Kac-Schwarz operator introduced in Eq.(3.93) of \cite{Agkm}, which is used to construct a Sato group element for $\tau_{KW}$. }

\subsection{The generalized Brezin–Gross–Witten tau-function}\label{S43}
The generalized BGW model $Z_N$ was first introduced in \cite{MMS} as the following deformed generalized Kontsevich matrix model
\begin{equation*}
	Z_{N}=\frac{\int [\d\Phi] \exp\left(\tr (\frac{\Lambda^2\Phi}{\h}+\frac{1}{\h\Phi}+(N-M)\ln \Phi) \right)}{\int [\d\Phi] \exp\left(\tr (\frac{1}{\h\Phi}+(N-M)\ln \Phi)\right)},
\end{equation*}
and its partition function $\tau_N(\h)$ is a tau-function of the MKP hierarchy  \cite{KMMM} with discrete time $N$. It is also a tau-function of KdV hierarchy for any $N\in\C$ under the Miwa parametrization $q_{2k+1}=\tr(\Lambda^{-2k-1})$. The case of $N=0$ is the original BGW tau-function. 

From the results in \cite{MMS} and \cite{Abgw}, we know that the following operators
\begin{align*}
	K_N&=\h^{-1} z+\frac{1}{2}D,\quad X_N=\h^{-2}z^2,\\
	P_N&=\frac{\h^2}{4}z^{-2}\left(D-\frac{1}{2}\right)^2+\h z^{-1}D-\frac{N^2\h^2}{4}z^{-2}
\end{align*}
are the Kac-Schwarz operator for $\tau_N(\h)$. The Virasoro constraints can be deduced using $K_N$ and $X_N$. In fact, the adjoint operator $z^{2m}K_N^*$ correspond to a set of Virasoro constraints $\{V_{2m}^{(N)}\}$ that annihilate and uniquely determine the tau-function $\tau_N$. That is,
\begin{equation*}
	V_{2m}^{(N)}\cdot\tau_N(\h)=0,
\end{equation*}
where
\begin{equation}\label{VforN}
	V_{2m}^{(N)}=\frac{1}{2}\widehat{L}_{2m}+\delta_{m,0}\left(\frac{1}{16}-\frac{N^2}{4}\right)-\h^{-1}(2m+1)\frac{\partial}{\partial q_{2m+1}}.
\end{equation}
Furthermore, the adjoint operator $P_N^*$ is in the form
\begin{equation*}	
	P_N^*=\frac{\h^2}{4}z^{-2}\left(D-\frac{1}{2}\right)^2-\h z^{-1}D-\frac{N^2\h^2}{4}z^{-2},
\end{equation*}
and the corresponding $W_{1+\infty}$ operator is
\begin{equation}
	W_{P_N^*}=-\frac{\h^2}{4}M_{-2}+\h L_{-1}-\frac{\h^2}{4}\left(\frac{1}{4}-N^2\right)q_2.
\end{equation}
The bosonic operator $\widehat{W}_{P_N^*}$ is a $W_{1+\infty}$ constraint of $\tau_N(\h)$, and it also uniquely determines the KdV tau-function $\tau_N(\h)$ (independent of $q_{2k}$) \cite{Abgw}:
\begin{equation*}
	\frac{\partial}{\partial q_{2k}}\cdot \tau_N=0 \quad\mbox{and}\quad\widehat{W}_{P_N^*}\cdot \tau_N=0.
\end{equation*}
\vspace{10pt}
\noindent

\begin{proof}[of Theorem \ref{WN}]
For $K_N^*=\h^{-1} z-\frac{1}{2}D$, we consider the operator
\begin{align*}
	P_N^*K_N^*&=-D-1+\frac{3\h}{4}z^{-1}D^2+\frac{\h}{4}z^{-1}D+\frac{\h}{4}\left(\frac{1}{4}-N^2\right)z^{-1}\\
	&\quad\quad-\frac{\h^2}{8}z^{-2}D^3+\frac{\h^2}{8}z^{-2}D^2-\frac{\h^2}{8}\left(\frac{1}{4}-N^2\right)z^{-2}D.
\end{align*}
From Eq.\eqref{nodes1} and \eqref{nodes2}, we have
\begin{align*}
	&W_{-z^{-1}D}=L_{-1},\quad W_{-z^{-1}D^2}=M_{-1},\\
	&W_{-z^{-2}D}=L_{-2}+\frac{1}{2}\alpha_{-2},\quad W_{-z^{-2}D^2}=M_{-2}+L_{-2}+\frac{1}{2}\alpha_{-2},\\
	&W_{-z^{-2}D^3}=Q_{-2}+\frac{3}{2}M_{-2}+L_{-2}+\frac{1}{2}\alpha_{-2}.
\end{align*}
Therefore,
\begin{align*}
	W_{P_N^*K_N^*}&=-D-1-\frac{3\h}{4}M_{-1}-\frac{\h}{4}L_{-1}-\frac{\h}{4}\left(\frac{1}{4}-N^2\right)\alpha_{-1}\\
	&\quad\quad+\frac{\h^2}{8}Q_{-2}+\frac{\h^2}{16}M_{-2}+\frac{\h^2}{8}\left(\frac{1}{4}-N^2\right)\left(L_{-2}+\frac{1}{2}\alpha_{-2}\right).
\end{align*}
Let
\begin{align*}
	W_{R_N}&=-D-W_{P_N^*K_N^*}-\frac{1}{4}W_{P_N^*}-1\\
	&=\frac{3\h}{4}M_{-1}+\frac{\h}{4}\left(\frac{1}{4}-N^2\right)\alpha_{-1}-
\frac{\h^2}{8}Q_{-2}-\frac{\h^2}{8}\left(\frac{1}{4}-N^2\right)L_{-2}.
\end{align*}
Then, Theorem \ref{WN} follows from Eq.\eqref{OEW}.
\end{proof}

\section{Reduction to the cut-and-join representation}\label{S5}

In order to simplify the expressions of our equations in the following context, we will use the notation $\Wa_i$ to represent $i\partial q_i$. Let us denote the set of positive odd integers to be $\Z_+^o$. We will also consider the operators $\widehat{\L}_n, \widehat{\M}_n$ and $\widehat{\Q}_n$ obtained by removing all the terms in $\widehat{L}_n, \widehat{M}_n$ and $\widehat{Q}_n$ containing even variables $q_{2k}$ and $\Wa_{2k}$, respectively. For example, in our case, we can write the Virasoro operators $\widehat{L}_{-2m-2}$ and $\widehat{L}_{2m}$ as
\begin{align*}
	\widehat{L}_{-2m-2}&=\sum_{i>0}q_{i+2m+2}\Wa_i+\frac{1}{2}\sum_{0<i<2m-2}q_iq_{2m+2-i},\\
	\widehat{L}_{2m}&=\sum_{j>0}q_j\Wa_{j+2m}+\frac{1}{2}\sum_{0<j<2m}\Wa_{j}\Wa_{2m-j},
\end{align*} 
and the operators $\widehat{\L}_{-2m-2}$ and $\widehat{\L}_{2m}$ as
\begin{align*}
	\widehat{\L}_{-2m-2}&=\sum_{i\in\Z_+^o}q_{i+2m+2}\Wa_i+\frac{1}{2}\sum_{\substack{i\in\Z_+^o\\0<i<2m-2}}q_iq_{2m+2-i},\\
	\widehat{\L}_{2m}&=\sum_{j\in\Z_+^o}q_j\Wa_{j+2m}+\frac{1}{2}\sum_{\substack{j\in\Z_+^o\\0<j<2m}}\Wa_{j}\Wa_{2m-j}.
\end{align*} 

The Kontsevich-Witten tau-function $\tau_{KW}(\h)$ has a very simple cut-and-join operator representation \cite{A1},
\begin{equation}\label{CJKW}
	\tau_{KW}(\h)=e^{\h^3 \widetilde{W}_{1}}\cdot 1,
\end{equation}
where
\begin{align}\label{CJKWo}
	\widetilde{W}_{1}&=\frac{1}{3}\sum_{k=0}^{\infty}q_{2k+1}\left(\widehat{\L}_{2k-2}+\frac{1}{8}\delta_{k,1}\right)\nonumber\\
	&=\frac{1}{3}\sum_{\substack{i,j\in\Z_+^o\\i+j>3}}^{\infty}q_{i}q_{j}\Wa_{i+j-3}+\frac{1}{6}\sum_{i,j\in\Z_+^o}^{\infty}q_{i+j+3}\Wa_i\Wa_j+\frac{1}{6}q_1^3+\frac{1}{24}q_3.
\end{align}
It is obtained by using the grading operator $\widehat{L}_0$ (or $\widehat{\L}_0$). In fact, all the Virasoro constraints for $\tau_{KW}(\h)$ can be packed into a single equation 
\begin{equation}\label{L0KW}
	\left(\widehat{\L}_0-3\h^3\widetilde{W}_{1}\right)\cdot \tau_{KW}(\h)=0,
\end{equation}
which is equivalent to 
\begin{equation*}
	\left(\partial_{\h}-3\h^2\widetilde{W}_{1}\right)\cdot \tau_{KW}(\h)=0.
\end{equation*}
The ordered exponential $\OE[3\h^2\widetilde{W}_{1}]$ is nothing but the exponential operator $\exp(\h^3 \widetilde{W}_{1})$.

Similarly, as introduced in \cite{Abgw}, the generalized BGW tau-function also has a cut-and-join representation: 
\begin{equation*}
	\tau_N(\h)=e^{\h \widetilde{W}_{N}}\cdot 1,
\end{equation*}
where
\begin{align}\label{CJNo}
	\widetilde{W}_{N}&=\sum_{k=0}^{\infty}q_{2k+1}\left\{\frac{1}{2}\widehat{\L}_{2k}+\delta_{k,0}\left(\frac{1}{16}-\frac{N^2}{4}\right)\right\}\nonumber\\
	&=\frac{1}{2}\sum_{i,j\in\Z_+^o}q_iq_j\Wa_{i+j-1}+\frac{1}{4}\sum_{i,j\in\Z_+^o}q_{i+j+1}\Wa_i\Wa_j+\left(\frac{1}{16}-\frac{N^2}{4}\right) q_1.
\end{align}
This is equivalent to $\left(\partial_{\h}-\widetilde{W}_{N}\right)\cdot \tau_N(\h)=0$ and
\begin{equation}\label{L0N}
	\left(\widehat{\L}_0-\h\widetilde{W}_{N}\right)\cdot \tau_N(\h)=0.
\end{equation}
The operator $\widetilde{W}_{N}$ actually corresponds to a BKP symmetry operator for $\tau_N$ (\cite{Abkp},\cite{Agbgw}). 
\vspace{10pt}
\noindent

\begin{proof}[of Corollary \ref{CJ}]
Let us set $\h=1$ in this proof for simplicity. We first give a proof of the reduction from Eq.\eqref{WNs} to Eq.\eqref{L0N} as follows. 
	
Since $\tau_N$ is independent of even variables $q_{2k}$, we can conclude from Eq.\eqref{WNs} that
\begin{equation}\label{L0No}
  \left\{\widehat{\L}_0-\frac{3}{4}\widehat{\M}_{-1}-\left(\frac{1}{16}-\frac{N^2}{4}\right) q_1+\frac{1}{8}\widehat{\Q}_{-2}+\frac{1}{8}\left(\frac{1}{4}-N^2\right)\widehat{\L}_{-2}\right\}\cdot\tau_N=0.
\end{equation}
Using Eq.\eqref{hatQ}, we can see that the expression of $\widehat{Q}_{-2}$ is
\begin{align*}
	\widehat{Q}_{-2}&=\sum_{i,j,k>0}q_iq_jq_k\Wa_{i+j+k-2}+\sum_{i,j,k>0}q_{i+j+k+2}\Wa_i\Wa_j\Wa_k\nonumber\\
	&+\frac{3}{2}\sum_{\substack{i,j,k>0\\i+j-k>2}}q_iq_j\Wa_k\Wa_{i+j-k-2}+\frac{1}{2}\sum_{i>0}(i+1)^2q_{i+2}\Wa_i.
\end{align*}
Therefore, the operator $\widehat{\mathcal{Q}}_{-2}$ would be
\begin{align*} 
	\widehat{\mathcal{Q}}_{-2}&=\sum_{\substack{i,j,k\in\Z_+^o\\i+j<k+2}}q_{i}q_{j}q_{k+2-i-j}\Wa_{k}+\sum_{i,j,k\in\Z_+^o}q_{i+j+k+2}\Wa_i\Wa_j\Wa_k\nonumber\\
	&\quad+\frac{3}{2}\sum_{\substack{i,j,k\in\Z_+^o\\i+j>k+2}}q_{i}q_{j}\Wa_k\Wa_{i+j-k-2}+2\sum_{n\geq 0}(n+1)^2q_{2n+3}\Wa_{2n+1}.
\end{align*}
Let us consider the summation
\begin{equation*}
	\sum_{m\geq 0}\widehat{\L}_{-2m-2}\widehat{\L}_{2m},
\end{equation*}
where
\begin{equation*}
	\widehat{\L}_{-2}\widehat{\L}_0=\sum_{i,j\in\Z_+^o}q_{i+2}q_j\Wa_i\Wa_j+\sum_{j\in\Z_+^o}jq_{j+2}\Wa_j+\frac{1}{2}\sum_{j\in\Z_+^o}q_1^2q_j\Wa_j.
\end{equation*}
and, for $m\geq 1$,
\begin{align*}
	&\widehat{\L}_{-2m-2}\widehat{\L}_{2m}\\
	=&\frac{1}{2}\sum_{\substack{i,j\in\Z_+^o\\0<i<2m+2}}q_{i}q_{2m+2-i}q_{j}\Wa_{j+2m}+\frac{1}{2}\sum_{\substack{i,j\in\Z_+^o\\0<j<2m}}q_{i+2m+2}\Wa_i\Wa_{j}\Wa_{2m-j}\\
	&\quad+\sum_{i,j\in\Z_+^o}q_{i+2m+2}q_{j}\Wa_i\Wa_{j+2m}+\sum_{j\in\Z_+^o}jq_{j+2m+2}\Wa_{j+2m}+\frac{1}{4}\sum_{\substack{i,j\in\Z_+^o\\0<i<2m+2\\0<j<2m}}q_{i}q_{2m+2-i}\Wa_{j}\Wa_{2m-j}.
\end{align*}
Observe that
\begin{align*}
	\sum_{m\geq 0}\sum_{i,j\in\Z_+^o}q_{i+2m+2}q_{j}\Wa_i\Wa_{j+2m}&=\frac{1}{2}\sum_{\substack{i,j,k\in\Z_+^o\\i+j>k+2}}q_{i}q_{j}\Wa_k\Wa_{i+j-k-2};\\
\sum_{m\geq 0}\sum_{\substack{i,j\in\Z_+^o\\0<i<2m+2\\0<j<2m}}q_{i}q_{2m+2-i}\Wa_{j}\Wa_{2m-j}&=\sum_{\substack{i,j,k\in\Z_+^o\\i+j>k+2}}q_{i}q_{j}\Wa_k\Wa_{i+j-k-2};\\
	\sum_{m\geq 0}\sum_{j\in\Z_+^o}jq_{j+2m+2}\Wa_{j+2m}&=\sum_{n\geq 0}q_{2n+3}\Wa_{2n+1}\sum_{m=0}^{n}\left(2n-2m+1\right)\\
	&=\sum_{n\geq 0}(n+1)^2q_{2n+3}\Wa_{2n+1}.
\end{align*}
This shows that
\begin{equation*}
\widehat{\mathcal{Q}}_{-2}=2\sum_{m\geq 0}\widehat{\L}_{-2m-2}\widehat{\L}_{2m}.
\end{equation*}
Since
\begin{equation*}
	\sum_{m\geq 0}\widehat{\L}_{-2m-2}\Wa_{2m+1}=\widehat{\M}_{-1}-\frac{1}{2}\sum_{i,j\in\Z_+^o}q_iq_j\Wa_{i+j-1},
\end{equation*}
where
\begin{equation*}
	\widehat{\M}_{-1}=\sum_{i,j\in\Z_+^o}q_{i+j+1}\Wa_i\Wa_j+\sum_{i,j\in\Z_+^o}q_iq_j\Wa_{i+j-1},
\end{equation*}
using the Virasoro constraints \eqref{VforN}, we can conclude that
\begin{align}
	\widehat{\mathcal{Q}}_{-2}\cdot\tau_N&=2\sum_{m\geq 0}\widehat{\L}_{-2m-2}\widehat{\L}_{2m}\cdot\tau_N\nonumber\\
	&=\sum_{m\geq 0}\widehat{\L}_{-2m-2}\left\{4\Wa_{2m+1}-\delta_{m,0}\left(\frac{1}{4}-N^2\right)\right\}\cdot\tau_N\nonumber\\
	&=\left\{4\widehat{\M}_{-1}-2\sum_{i,j\in\Z_+^o}q_iq_j\Wa_{i+j-1}-\left(\frac{1}{4}-N^2\right)L_{-2}\right\}\cdot\tau_N\label{Q2N}
\end{align}
On the other hand, we can re-write $\widetilde{W}_{N}$ in Eq.\eqref{CJNo} as
\begin{equation}\label{WN-1}
	\widetilde{W}_{N}=\frac{1}{4}\widehat{\M}_{-1}+\frac{1}{4}\sum_{i,j\in\Z_+^o}q_iq_j\Wa_{i+j-1}+\left(\frac{1}{16}-\frac{N^2}{4}\right) q_1.
\end{equation}
After combining Eq.\eqref{L0No}, \eqref{Q2N} and \eqref{WN-1}, we can obtain Eq.\eqref{L0N}. 

The reduction from Eq.\eqref{W1s} to Eq.\eqref{L0KW} for Kontsevich-Witten tau-function is similar. First, from Eq.\eqref{W1s}, we have
\begin{equation}\label{L0KWo}
	\left( \widehat{\L}_0-\frac{3}{2}\widehat{\M}_{-3}-\frac{1}{8}q_3+\frac{1}{2}\widehat{\Q}_{-6}+\frac{21}{40}\widehat{\L}_6\right)\cdot \tau_{KW}=0.
\end{equation}
The operator $\widehat{\Q}_{-6}$ is in the form
\begin{align*}
	\widehat{\Q}_{-6}&=q_1^3q_3+\sum_{\substack{i,j,k\in\Z_+^o\\i+j+k>6}}q_iq_jq_k\Wa_{i+j+k-6}+\sum_{i,j,k\in\Z_+^o}q_{i+j+k+6}\Wa_i\Wa_j\Wa_k\nonumber\\
	&+\frac{3}{2}\sum_{\substack{i,j,k\in\Z_+^o\\i+j-k>6}}q_iq_j\Wa_k\Wa_{i+j-k-6}+\frac{1}{2}\sum_{i\in\Z_+^o}(i+1)(i+5)q_{i+6}\Wa_i-q_3^2+\frac{6}{5}\widehat{\L}_{-6}.
\end{align*}
Note that the quadratic term in
\begin{equation*}
	\frac{1}{2}\widehat{\mathcal{Q}}_{-6}+\frac{21}{40}\widehat{\L}_{-6}
\end{equation*} 
can be expressed as
\begin{align*}
	&\frac{1}{4}\sum_{i\in\Z_+^o}(i+1)(i+5)q_{i+6}\Wa_i-\frac{1}{2}q_3^2+\frac{9}{8}\widehat{\L}_{-6}\\
	=&\sum_{n=0}^{\infty}(n+2)^2q_{2n+7}\Wa_{2n+1}-\sum_{n=0}^{\infty}q_{2n+7}\Wa_{2n+1}-\frac{1}{2}q_3^2+\frac{9}{8}\widehat{\L}_{-6},
\end{align*}
and the quadratic term in 
\begin{equation*}
	\sum_{m\geq -1}\widehat{\L}_{-2m-6}\widehat{\L}_{2m}
\end{equation*}
is 
\begin{align*}
	&q_1q_5+\sum_{i\in\Z_+^o}(i+2)q_{i+6}\Wa_{i}+\sum_{m\geq 0}\sum_{j\in\Z_+^o}jq_{j+2m+6}\Wa_{j+2m}\\
	=&q_1q_5+\sum_{n\geq 0}q_{2n+7}\Wa_{2n+1}\sum_{m=-1}^{n}\left(2n-2m+1\right)\\
	=&q_1q_5+\sum_{n\geq 0}(n+2)^2q_{2n+7}\Wa_{2n+1}.
\end{align*}
Therefore, we can obtain 
\begin{align*}
	&\frac{1}{2}\widehat{\mathcal{Q}}_{-6}+\frac{21}{40}\widehat{\L}_{-6}\\
	=&\sum_{m\geq -1}\widehat{\L}_{-2m-6}\widehat{\L}_{2m}-q_1q_5-\sum_{n=0}^{\infty}q_{2n+7}\Wa_{2n+1}-\frac{1}{2}q_3^2+\frac{9}{8}\widehat{\L}_{-6}\\
	=&\sum_{m\geq -1}\widehat{\L}_{-2m-6}\widehat{\L}_{2m}+\frac{1}{8}\widehat{\L}_{-6}
\end{align*}
Since
\begin{equation*}
	\sum_{m\geq -1}\widehat{\L}_{-2m-6}\Wa_{2m+3}=\widehat{\M}_{-3}-\frac{1}{2}\sum_{\substack{i,j\in\Z_+^o\\i+j>3}}q_iq_j\Wa_{i+j-3}-\frac{1}{3}q_1^3,
\end{equation*}
where
\begin{equation*}
	\widehat{\M}_{-3}=\sum_{i,j\in\Z_+^o}q_{i+j+3}\Wa_i\Wa_j+\sum_{\substack{i,j\in\Z_+^o\\i+j>3}}q_iq_j\Wa_{i+j-3}+\frac{1}{3}q_1^3,
\end{equation*}
using the Virasoro constraints \eqref{VforK}, we can conclude that
\begin{align}\label{Q6kw}
	\left(\frac{1}{2}\widehat{\mathcal{Q}}_{-6}+\frac{21}{40}\widehat{\L}_{-6}\right)\cdot\tau_{KW}
	=&\left(\sum_{m\geq -1}\widehat{\L}_{-2m-6}\widehat{\L}_{2m}+\frac{1}{8}\widehat{\L}_{-6} \right)\cdot\tau_{KW}\nonumber\\
	=&\left(\widehat{\M}_{-3}-\frac{1}{2}\sum_{\substack{i,j\in\Z_+^o\\i+j>3}}q_iq_j\Wa_{i+j-3}-\frac{1}{3}q_1^3\right)\cdot\tau_{KW}.
\end{align}
On the other hand, from Eq.\eqref{CJKWo}, we can see that
\begin{align}\label{WKW1}
	3\widetilde{W}_{1}&=\sum_{\substack{i,j\in\Z_+^o\\i+j>3}}^{\infty}q_{i}q_{j}\Wa_{i+j-3}+\frac{1}{2}\sum_{i,j\in\Z_+^o}^{\infty}q_{i+j+3}\Wa_i\Wa_j+\frac{1}{2}q_1^3+\frac{1}{8}q_3\nonumber\\
	&=\frac{1}{2}\widehat{\M}_{-3}+\frac{1}{2}\sum_{\substack{i,j\in\Z_+^o\\i+j>3}}^{\infty}q_{i}q_{j}\Wa_{i+j-3}+\frac{1}{3}q_1^3+\frac{1}{8}q_3.
\end{align}
Therefore, Eq.\eqref{L0KW} follows from Eq.\eqref{L0KWo}, \eqref{Q6kw} and \eqref{WKW1}. This completes the proof.
\end{proof}

\section*{Acknowledgement}
The author would like to thank Alexander Alexandrov for valuable suggestions. This work was supported by the National Natural Science Foundation of China (Grant No.11701587).

\vspace{10pt} \noindent
\\
\footnotesize{\sc gehao wang }\\
Department of Mathematics,\\
College of Information Science and Technology/College of Cyberspace Security,\\
Jinan University, Guangzhou, China. \\
\footnotesize{E-mail address:  gehao\_wang@hotmail.com}

\end{document}